\documentclass{article}
\usepackage{geometry}                
\usepackage{amsmath,amssymb,amsthm,times,url}
\usepackage[sort&compress]{natbib}
\usepackage{graphicx}
\usepackage{epstopdf}
\def\today{\number\day\ \ifcase\month\or
 January\or February\or March\or April\or May\or June\or
 July\or August\or September\or October\or November\or December\fi
 \space \number\year}
\def\ds{\displaystyle}
\def\hide#1{}
\def\figs#1{}

\def\ifrac#1#2{{#1}/{#2}}

\def\part#1#2{\boldsymbol{\la}({#1,#2})}

\def\d{{\rm d}}
\def\e{{\rm e}}
\def\i{\ifmmode{\rm i}\else\char"10\fi}
\def\D{{\rm D}}

\def\p{Pain\-lev\'e}
\def\sch{Schr\"odinger}

\def\bq{Boussinesq}
\def\AM{Adler--Moser}

\def\peq{\p\ equation}
\def\peqs{\p\ equations}

\def\PI{\mbox{\rm P$_{\rm I}$}}
\def\PII{\mbox{\rm P$_{\rm II}$}}

\def\PIV{\mbox{\rm P$_{\rm IV}$}}

\def\PVI{\mbox{\rm P$_{\rm VI}$}}

\def\tfrac#1#2{{\textstyle\frac{#1}{#2}}}
\def\cdot{{\scriptstyle\,\bullet\,}}

\def\odes{ordinary differential equations}

\def\pdes{partial differential equations}
\newcommand{\deriv}[3][]{\frac{\d^{#1}{#2}}{{\d{#3}}^{#1}}}

\newcommand{\ideriv}[3][]{{\d^{#1}{#2}}/{{\d{#3}}^{#1}}}

\def\D{{\rm D}}

\def\bfk{\boldsymbol{\kappa}}

\def\a{\alpha}
\def\b{\beta}

\def\la{\lambda}
\def\k{\kappa}
\def\ph{\varphi}
\def\z{\zeta}
\def\th{\theta}

\def\Th{\Theta}
\def\Tht{\widetilde{\Th}}
\def\part#1#2{\boldsymbol{\la}(#1,#2)}

\def\GG{\Omega}

\def\VY{Yablonskii--Vorob'ev polynomials}
\def\Ok{Okamoto polynomials}

\newcommand{\Com}{\mathbb{C}}

\newcommand{\NatNumber}{\mathbb{N}}
\newcommand{\Integer}{\mathbb{Z}}

\def\Z{\Integer}
\def\N{\NatNumber}

\def\G#1{\Gamma_{\!#1}}
\def\zz{z^*}
\def\ssum{\sideset{}{'}\sum}
\def\bfkappa{\boldsymbol{\kappa}}

\newtheorem{theorem}{Theorem}
\newtheorem{definition}[theorem]{Definition}

\numberwithin{equation}{section}
\numberwithin{table}{section}
\numberwithin{figure}{section}
\numberwithin{theorem}{section}

\newcommand{\Vector}[1]{\ifcat#1a\mathbf{#1}\else\boldsymbol{#1}\fi}
\newcommand{\Matrix}[1]{\ifcat#1a\mathbf{#1}\else\boldsymbol{#1}\fi}

\begin{document}

\title{Vortices and Polynomials}

\author{Peter A.~Clarkson\\ Institute 
of Mathematics, Statistics and Actuarial
Science,\\ University of Kent, Canterbury, CT2 7NF, UK\\
\texttt{P.A.Clarkson@kent.ac.uk}}

\maketitle 
\thispagestyle{empty}

\begin{abstract}
The relationship between point vortex dynamics and the properties of polynomials with roots at the vortex positions is discussed. Classical polynomials, such as the Hermite polynomials, have roots that describe the equilibria of identical vortices on the line. Stationary and uniformly translating vortex configurations with vortices of the same strength but positive or negative orientation are given by the zeros of the \AM\ polynomials, which arise in the description of rational solutions of the Korteweg-de Vries equation. For quadrupole background flow, vortex configurations are given by the zeros of polynomials expressed as wronskians of Hermite polynomials. Further new solutions are found in this case using the special polynomials arising the in the description of rational solutions of the fourth \p\ equation.
\end{abstract}

\thispagestyle{plain}

\section{Introduction}
Vortices are some of the most visually appealing phenomena of fluid mechanics. Examples include the bathtub vortex, vortices resulting from flow past obstacles, wingtip vortices, as well as hurricanes and tornadoes. Vortical flows play an essential role in many applications and hence it is no surprise that vortex dynamics is an important
sub-topic in fluid mechanics. A polynomial is a basic mathematical structure, perhaps the simplest and most fundamental. There are many situations where approximations of mathematical functions are made using polynomials, for example in numerical methods. In this paper we are concerned with the relationship between vortex dynamics and the roots of some special polynomials.

The point vortex equations may be thought of as a discretization of the equations for two-dimensional flow, the \textit{Euler equations}, which is useful for analytical and numerical approximations. Considering the flow plane of the two-dimensional fluid to be the complex plane, i.e.\ the point $(x,y)$ is viewed as the complex number $z=x+\i y$, one associates a set of point vortices with a polynomial that has roots at the locations of the vortices and uses this representation to study the vortex configurations.

In two recent papers, Aref \citep{refAref07a,refAref07b} reviews the connection between point vortex dynamics and polynomials with roots at the positions of the vortices. For stationary vortex configurations the following results have been established: \begin{enumerate}\item $N$ identical vortices on a line are in equilibrium if and only if they are situated at the roots of the classical $N$th Hermite polynomial, which are related to the \textit{Stieltjes relations} \citep{refStieltjes85a,refStieltjes85b}, see equation (\ref{eq2}); \item $N$ identical vortices on a circle are in equilibrium if and only if they are situated at the vertices of a regular $N$-polygon \citep{refLewis}; \item $\frac12n(n+1)$ positive and $\frac12n(n-1)$ negative vortices are in equilibrium if and only if they are situated at the roots of the \textit{Adler--Moser polynomials} \citep{refANSTV,refBartman,refKC87}
(see also \S\ref{ssec:AM} and \S\ref{sec:svp});
\item $N$ positive and $N$ negative vortices form a uniformly translating equilibrium if and only if
the vortices of one sign are situated at the roots of an Adler--Moser polynomial and the
vortices of opposite sign are then at the roots of a polynomial of the same degree derived
from the Adler--Moser polynomial \citep{refANSTV,refBartman,refKC87} (see also \S\ref{sec:tvp}). Further, such equilibria are only possible when $N$ is a triangular number.  \end{enumerate}
Aref \citep{refAref07a,refAref07b} remarks that the relationship between the theory of stationary equilibria and rational solutions of the Korteweg-de Vries (KdV) equation
\begin{equation} u_t + 6uu_x + u_{xxx} =0,\label{eq:kdv} \end{equation}
one of the original contexts in which the Adler--Moser polynomials arose, is ``quite unexpected and very beautiful".

There has been considerable interest in completely integrable \pdes\ solvable by inverse scattering, the \textit{soliton equations}, since the discovery in 1967
by Gardner, Greene, Kruskal, and Miura \citep{refGGKM} of the method for
solving the initial value problem for the KdV equation (\ref{eq:kdv}).
During the past thirty years or so there have been several studies of rational solutions for the soliton equations, applications of which include the description of explode-decay waves \citep{refHN85} and vortex solutions of
the complex sine-Gordon equation \citep{refBP98,refOB05}. Airault, McKean, and Moser \citep{refAMM77} studied the motion of the poles of rational solutions of the KdV equation (\ref{eq:kdv}) and related the motion to an integrable many-body problem, the Calogero-Moser system with constraints. Subsequently Adler and Moser \citep{refAdMoser} expressed the rational solutions of the KdV equation in terms of the \AM\ polynomials (see \S\ref{ssec:AM} below). Ablowitz and Satsuma \citep{refASat} derived rational solutions of the KdV equation (\ref{eq:kdv}) by finding a long-wave limit of the known $N$-soliton solutions of these equations. 

The \textit{\peqs} (\PI--\PVI) are six nonlinear \odes\ discovered by \p, Gambier and their colleagues around the beginning of the 20th century. Their solutions define new transcendental functions as they are not expressible in terms of previously known functions such as elementary and elliptic functions or in terms of solutions of linear \odes\ and can be thought of as nonlinear analogues of the classical special functions \citep{refPAC05review,refFIKN,refGLS02,refUmemura98}. It is well known that \PII--\PVI\ have rational solutions, algebraic solutions, and solutions expressed in terms of the classical special functions (cf.\ \citep{refPAC05review,refGLS02}).
Ablowitz and Segur \citep{refAS77} demonstrated a close relationship between the soliton and \peqs. For example, the second \peq\ (\PII)
\begin{equation}\label{eq:PII} \deriv[2]{w}z = 2w^3 + z w + \a,\end{equation} 
where $\a$ is an arbitrary constant, arises as a scaling reduction of the KdV equation (\ref{eq:kdv}) and the fourth \peq\ (\PIV)
\begin{equation}\label{eq:PIV} w\deriv[2]{w}z = \frac12\left(\deriv{w}z\right)^{\!2} +
\frac{3}{2}w^4 + 4z w^3 + 2(z^2 - \a)w^2 +{\b},\end{equation}
where $\a$ and $\b$ are arbitrary constants, arises as scaling reductions of the \bq\ equation 
\begin{equation}\label{eq:bq}
u_{tt}+(u^2)_{xx}\pm u_{xxxx}=0,
\end{equation} 
and the nonlinear \sch\ (NLS) equation
\begin{equation}\label{eq:nls}
\mbox{i} q_t + q_{xx} \pm 2|q|^2q=0.
\end{equation}

Vorob'ev and Yablonskii \citep{refVor,refYab} expressed the rational solutions of \PII\ (\ref{eq:PII}) in terms of polynomials, now known as the \emph{\VY} (see \S\ref{ssec:PII} below), 
which are special cases of the {Adler--Moser polynomials}. 
Okamoto \citep{refOkamotoiii} derived analogous polynomials, now known as the \emph{\Ok}, related to some of the rational solutions of \PIV\ (\ref{eq:PIV}). Okamoto's results were generalized by Noumi and Yamada \citep{refNY99i} who expressed rational solutions of \PIV\ (\ref{eq:PIV}) in terms of two sets of polynomials, the \emph{generalized Hermite polynomials} and the \emph{generalized Okamoto polynomials}\  (see \S\ref{ssec:PIV} below). 

This paper is organized as follows. In \S\ref{sec2} we describe the special polynomials associated with rational solutions of the \p\ and soliton equations. In \S\ref{sec:vortex} we discuss several examples of equilibrium vortex configurations for vortices of the same strength but positive or negative orientation. In \S\ref{ssec:3.2} stationary vortex configurations and uniformly translating vortex configurations are shown to be given by the zeros of the \AM\ polynomials and related to eigenstates of the \sch\ equation for rational potentials decaying at infinity, which relates to work of Duistermaat and Gr\"unbaum \citep{refDG86}. In \S\ref{ssec:3.3}, for quadrupole background flow, vortex configurations are shown to be given by the zeros of polynomials expressed as wronskians of Hermite polynomials and related to eigenstates of the \sch\ equation for rational potentials with quadratic growth at infinity which relates to work of Oblomkov \citep{refOblomkov}, Veselov \citep{refVeselov}, and Loutsenko \citep{refLout03}. Further using results on the generalized Hermite and generalized Okamoto polynomials associated with rational solutions of \PIV\ (\ref{eq:PIV}) it is shown that there are more solutions in the case of quadrupole background flow than might be anticipated from previous work. Finally in \S\ref{sec4} we discuss our results.

\section{\label{sec2}Special polynomials associated with the \p\ and soliton equations}

\subsection{\label{ssec:PII}The \VY}
Rational solutions of \PII\ (\ref{eq:PII}), with $\alpha=n\in\Integer$, can be expressed in terms of special polynomials now known as the {\it\VY\/}, which were introduced by Vorob'ev \citep{refVor} and Yablonskii \citep{refYab} and are defined in Definition \ref{defn:21}.
Kajiwara and Ohta \citep{refKO96} expressed rational solutions of \PII\ (\ref{eq:PII}) in terms of Schur functions by expressing them in the form of determinants, which is how they are defined below (see also \citep{refCM03,refKMi,refUmemura01}); we remark that Flaschka and Newell \citep{refFN}, following the earlier
work of Airault \citep{refAirault}, expressed the rational solutions of \PII\ (\ref{eq:PII}) in terms of determinants.

\begin{definition}{\label{defn:21}\rm Consider the polynomials $\ph_n(z)$ defined by
\begin{equation}\label{eq:2.1}\sum_{n=0}^\infty\ph_n(z)\la^n
=\exp\left(z\la-\tfrac43\la^3\right),\end{equation}
and then the \textit{Yablonskii--Vorob'ev\ polynomial}\ $Q_{n}(z)$ is defined by
\begin{equation}Q_{n}(z)=c_n\mathcal{W}(\ph_1,\ph_3,\ldots,\ph_{2n-1}),
\qquad c_n=\prod_{j=1}^{n}(2j+1)^{n-j},
\end{equation} where $\mathcal{W}(\ph_1,\ph_3,\ldots,\ph_{2n-1})$ is the usual
Wronskian.}\end{definition}

Examples of \VY\ and plots of the locations of their roots in the complex plane are given by Clarkson and Mansfield \citep{refCM03}, who showed that the roots have a very regular, approximately triangular structure. The term ``approximate" is used since the patterns are not exact triangles as the roots lie on arcs rather than straight lines. 

\subsection{\label{ssec:AM}The \AM\ polynomials}
Burchnall and Chaundy [3] determined what condition must be satisfied by two polynomials $P(z)$ and $Q(z)$ in
such that the indefinite integrals
$$\int\frac{P^2(z)}{Q^2(z)}\,\d z,\qquad\int\frac{Q^2(z)}{P^2(z)}\,\d z,$$
be rational, i.e.\ expressible without logarithms. Under the assumption that $P(z)$ and $Q(z)$ have
no common or repeated factors, the problem is reduced to finding polynomial solutions of the bilinear equation
\begin{equation}\label{eq26a}
Q\deriv[2]{P}{z}-2\deriv{P}{z}\deriv{Q}{z}+P\deriv[2]{P}{z}=0.\end{equation}

Airault, McKean, and Moser \citep{refAMM77} studied the motion of the poles of rational solutions of the KdV equation (\ref{eq:kdv}) and related the motion to an integrable many-body problem, the Calogero-Moser system, with the Hamiltonian
$$\mathcal{H}=\sum_{j=1}^Np_j^2+\sum_{j=1}^N\sum_{k=j+1}^N\frac{2}{(z_j-z_k)^2},$$
which are described by the following algebraic system
\begin{equation}\ssum_{k=1}^N\frac{1}{(z_j-z_k)^3}=0,\qquad j=1,2,\ldots,N,\label{eq:AMMsys}\end{equation}
where $\ds\ssum_{k=1}^{N}$ means $\ds\sum_{{\scriptstyle k=1,\scriptstyle k\not=j}}^{N}$.
A remarkable fact, first discovered by Airault, McKean, and Moser \cite{refAMM77}, is that the system (\ref{eq:AMMsys}) has no solutions unless $N=\tfrac12n(n+1)$, with $n\in\Z$, i.e.\ a triangular number, in which case the solutions depend on $n$ arbitrary complex parameters. Adler and Moser \citep{refAdMoser} constructed  the polynomials $\Th_n(z)=\prod_{k=1}^N(z-z_k)$, which are defined in Definition \ref{def:ampoly} below, whose roots satisfy (\ref{eq:AMMsys}) and so all solutions of  (\ref{eq:AMMsys}) can be obtained. These are constructed by Adler and Moser \citep{refAdMoser} using of the Darboux transformations of the operator $\ds\deriv[2]{}{z}$, following the work of Crum \citep{refCrum}, though many of the results appear in the paper by Burchnall and Chaundy \citep{refBC29}; see also \citep{refLout04}.

\begin{definition}{\label{def:ampoly}\rm Consider the polynomials $\th_n(z)$ defined by
\begin{equation}\label{eq:2.5}\sum_{n=0}^\infty\th_n(z)\la^n
=\exp\left(z\la-\sum_{j=2}^\infty\frac{\k_j\la^{2j-1}}{2j-1}\right),\end{equation}
with $\k_j$, $j=2,3,\ldots,$ arbitrary parameters, and then the \textit{\AM\ polynomial}\
$\Th_{n}(z;\bfkappa_n)$  is defined by
\begin{equation}\Th_{n}(z;\bfkappa_n)=c_n\mathcal{W}(\th_1,\th_3,\ldots,\th_{2n-1}),
\qquad c_n=\prod_{j=1}^{n}(2j+1)^{n-j},
\end{equation} 
where $\bfkappa_n=(\k_2,\k_3,\ldots,\k_n)$. 
The \textit{modified \AM\ polynomial}\ $\Tht_{n}(z;\mu,\bfk_{n})$ is defined by
\begin{equation}\Tht_{n}(z;\mu,\bfk_{n})=c_n\e^{-\mu z}\mathcal{W}(\th_1,\th_3,\ldots,\th_{2n-1},\e^{\mu z}),\end{equation}
with $\mu$ an arbitrary constant.}\end{definition}

We remark that it can be shown that \AM\ and modified \AM\ polynomials are related by
$$\Tht_{n}(z;\mu,{\bfk}_{n})=\Th_{n}\left(z-\ifrac1{\mu};\widetilde{\bfk}_{n}\right),\qquad
\widetilde{\bfk}_{n}=\left(\k_2+\ifrac1{\mu^{3}},\k_3+\ifrac1{\mu^{5}},\ldots,\k_n+\ifrac1{\mu^{2n-1}}\right).$$ 
Further, we note that the \VY\ $Q_n(z)$ are special cases of the \AM\ polynomials 
$\Th_{n}(z;{\bfk}_{n})$ with $\k_2=4$ and $\k_n=0$ for $n\geq3$.

As shown by Burchnall and Chaundy \citep{refBC29}, and later by Adler and Moser \citep{refAdMoser}, the \AM\ polynomials satisfy the bilinear equations 
\begin{align} \label{eq26}
&\deriv[2]{\Th_{n+1}}{z}\Th_n-2\deriv{\Th_{n+1}}{z}\deriv{\Th_{n}}{z}+\Th_{n+1}\deriv[2]{\Th_{n}}{z}=0,\\
&\deriv{\Th_{n+1}}{z}\Th_{n-1}-\Th_{n+1}\deriv{\Th_{n-1}}{z}=(2n+1)\Th_n^2.\label{eq27}
\end{align}
In fact equation (\ref{eq27}) is often used to generate the \AM\ polynomials, given $\Th_0(z)=1$ and $\Th_1(z)=z$, with the parameters $\k_j,j=2,3,\ldots,$ arising as constants of integration.

Bartman \citep{refBartman} showed that roots of the consecutive polynomials $\Th_{n-1}(z)$ and $\Th_{n}(z)$ are the equilibrium coordinates of $\tfrac12n(n-1)$ positive and $\tfrac12n(n+1)$ negative Coulomb charges, with values $\pm1$ respectively, on the plane interacting through a two-dimensional (logarithmic Coulomb) potential. Namely, the function
\[E=\sum_{j=1}^{n(n-1)/2}\sum_{k=j+1}^{n(n-1)/2} \ln|z_j-z_k|+ 
\sum_{j=1}^{n(n+1)/2}\sum_{k=j+1}^{n(n+1)/2} \ln|\zeta_j-\zeta_k|+
\sum_{j=1}^{n(n-1)/2}\sum_{k=1}^{n(n+1)/2} \ln|z_j-\zeta_k|,\]
has a critical point when $z_j$, $j=1,2,\ldots,\tfrac12n(n-1)$, and $\zeta_k$, $k=1,2,\ldots,\tfrac12n(n+1)$, are roots of $\Th_{n-1}(z)$ and $\Th_{n}(z)$ respectively. 

\subsection{\label{ssec:PIV}The generalized Hermite and generalized Okamoto polynomials}
In a comprehensive study of properties of solutions of \PIV\ (\ref{eq:PIV}), Okamoto
\citep{refOkamotoiii} introduced two sets of polynomials associated with
rational solutions of \PIV\ (\ref{eq:PIV}), analogous to the {Yablonskii--Vorob'ev polynomials} discussed in \S\ref{ssec:PII}. Noumi and Yamada \citep{refNY99i} generalized Okamoto's results and introduced the \emph{generalized Hermite polynomials}, which are defined in Definition \ref{genherm}, and the \emph{generalized \Ok}, which are defined in Definition \ref{genokpolytilde}; see also \citep{refPAC03piv,refPAC06cmft}. Kajiwara and Ohta \citep{refKO98} expressed the generalized Hermite and generalized \Ok\ in terms of Schur functions in the form of determinants, which is how they are defined below. 

\begin{definition}\label{genherm}{\rm The \textit{generalized Hermite polynomial}\ $H_{m,n}(z)$ is defined by
\begin{align}H_{m,n}(z)&=\mathcal{W}\left(H_{m}(z), H_{m+1}(z), \ldots,H_{m+n-1}(z)\right)
\equiv\mathcal{W}\left(\big\{H_{m+j}(z)\big\}_{j=0}^{n-1}\right),\qquad m,n\geq1, \end{align} 
with $H_{m,0}(z)=H_{0,n}(z)=1$ and $H_{k}(z)$ the $k$th Hermite polynomial,
and has degree $\mbox{deg}(H_{m,n}(z))=mn$.
}\end{definition}

Examples of generalized Hermite polynomials and plots of the locations of their roots in the complex plane are given by Clarkson \citep{refPAC03piv}; see also \citep{refPAC06cmft,refPAC06nls,refPACbqrats}. The roots take the form of $m\times n$ ``rectangles", which are only approximate rectangles since the roots lie on arcs rather than straight lines. 

From Lemma 5.7 in \citep{refNY99i} 
it follows that the generalized Hermite polynomials $H_{m,n}(z)$ satisfy the bilinear equations
\begin{subequations}\label{eq:bilinHmn}\begin{align}
&\left[\D_z^2-2z \D_z-2n\right]H_{m,n}\cdot H_{m+1,n}=0,\\
&\left[\D_z^2-2z \D_z+2m\right]H_{m,n+1}\cdot H_{m,n}=0,\\
&\left[\D_z^2-2z \D_z+2(m-n)\right]H_{m,n+1}\cdot H_{m+1,n}=0,
\end{align}\end{subequations}
with $\D_z$ the Hirota operator defined by
\begin{equation}\label{Hirop}\D_z^j(F\cdot G)(z)=\left.\deriv[j]{}{\zeta}\left\{
F(z+ \zeta)G(z-\zeta)\right\}\right|_{\zeta =0}. \end{equation}

The generalized Hermite polynomial $H_{m,n}(z)$ can be expressed as the multiple integral
\begin{align*}H_{m,n}(z) &= \frac{\pi^{m/2}\prod_{k=1}^mk!}{2^{m(m+2n-1)/2}}
\int_{-\infty}^{\infty}\!\!\!\begin{array}{c}{\cdots}\\[-10pt]{}_{n}\end{array}\!\!
\int_{-\infty}^{\infty}
\prod_{i=1}^{n}\prod_{j=i+1}^{n}(x_i-x_j)^2
\prod_{k=1}^n(z-x_k)^m \exp\left(-x_k^2\right)
\d x_1\,\d x_2\ldots\d x_n,\end{align*}
which arises in random matrix theory \citep{refBH00,refFW01, refKanzieper}. 
The generalized Hermite polynomials also arise in the theory of orthogonal polynomials \citep{refCF06}.

\begin{definition}\label{genokpolytilde}{\rm The \textit{generalized Okamoto polynomial} $\GG_{m,n}(z)$ is defined by
\begin{subequations}\begin{align}
\GG_{m,n}(z) &=\mathcal{W}(H_1(z),H_4(z),\ldots,H_{3m-2}(z),H_2(z),H_5(z),\ldots,H_{3n-1}(z)),\\
&\equiv\mathcal{W}\left(\big\{H_{3j+1}(z)\big\}_{j=0}^{m-1},\big\{H_{3k+2}(z)\big\}_{k=0}^{n-1}\right),
\qquad m,n\geq1,\nonumber\\
\GG_{m,0}(z) &=\mathcal{W}(H_1(z),H_4(z),\ldots,H_{3m-2}(z))
\equiv\mathcal{W}\left(\big\{H_{3j+1}(z)\big\}_{j=1}^{m-1}\right),\qquad m\geq1,\\
\GG_{0,n}(z) &=\mathcal{W}(H_2(z),H_5(z),\ldots,H_{3n-1}(z))
\equiv\mathcal{W}\left(\big\{H_{3k+2}(z)\big\}_{j=1}^{n-1}\right),\qquad n\geq1,
\end{align}\end{subequations}
with $\GG_{0,0}(z)=1$ and $H_{k}(z)$ the $k$th Hermite polynomial,
and has degree $\mbox{deg}(\GG_{m,n}(z))=m^2+n^2-mn+n$.
}\end{definition}

We note that the generalized Okamoto polynomial $\GG_{m,n}(z)$ defined here have been reindexed in comparison to the generalized Okamoto polynomial $Q_{m,n}(z)$ defined in \citep{refPAC03piv,refPAC06cmft} by setting $Q_{m,n}(z)=\GG_{m+n-1,n-1}(z)$ and $Q_{-m,-n}(z)=\GG_{n-1,m+n}(z)$, for $m,n\geq1$. The polynomials introduced by Okamoto \citep{refOkamotoiii} are given by $Q_{m}(z)=\GG_{m-1,0}(z)$ and $R_{m}(z)=\GG_{m,1}(z)$. Further the generalized Okamoto polynomial introduced by Noumi and Yamada \citep{refNY99i} is given by $\widetilde{Q}_{m,n}(z)=\GG_{m-1,n-1}(z)$.

Examples of generalized \Ok\ and plots of the locations of their roots in the complex plane are given by Clarkson \citep{refPAC03piv,refPAC06cmft}. The roots of the polynomial $Q_{m,n}(z)=\GG_{m+n-1,n-1}(z)$ with $m,n\geq1$ take the form of an $m\times n$ ``rectangle" with an ``equilateral triangle", which have either $\tfrac12m(m-1)$ or $\tfrac12n(n-1)$ roots, on each of its sides. The roots of the polynomial $Q_{-m,-n}(z)=\GG_{n-1,m+n}(z)$ with $m,n\geq1$ take the form of an $m\times n$ ``rectangle" with an ``equilateral triangle", which now have either $\tfrac12m(m+1)$ or $\tfrac12n(n+1)$ roots, on each of its sides. Again these are only approximate rectangles and equilateral triangles since the roots lie on arcs rather than straight lines. 

From Lemma 5.7 in \citep{refNY99i} (see also \citep{refLout03}), 
it can be shown that the generalized Okamoto polynomials $\GG_{m,n}(z)$ satisfy the bilinear equations
\begin{subequations}\label{eq:bilinQmn}\begin{align}
&\left[\D_z^2-2z \D_z+2(m-2n+2)\right]\GG_{m+1,n}\cdot \GG_{m,n}=0,\\
&\left[\D_z^2-2z \D_z-2(2m-n-1)\right]\GG_{m,n+1}\cdot \GG_{m,n}=0,\\
&\left[\D_z^2-2z \D_z-2(m+n+2)\right]\GG_{m,n}\cdot \GG_{m+1,n+1}=0,
\end{align}\end{subequations}
with $\D_z$ the Hirota operator (\ref{Hirop}).

The symmetric fourth \p\ (s\PIV) hierarchy introduced by Noumi and Yamada \citep{refNY98iii}, which has the affine Weyl group symmetry  $A^{(1)}_{2n}$, for $n\in \NatNumber $, is given by
\begin{subequations}\label{eq:sP4h}\begin{equation}
\deriv{\ph_j}{z}+\ph_j\left(\sum_{r=1}^{n}\ph_{j+2r-1}
-\sum_{r=1}^{n}\ph_{j+2r}\right)+2\mu_j=0,\qquad j=0,1,\ldots,
2n,\end{equation} subject to the constraints
\begin{equation} \sum_{j=0}^{2n} \ph_j=-2z, \qquad
\sum_{j=0}^{2n}\mu_j=1,
\end{equation}\end{subequations} where 
$\mu_j$, for $j=0,1,\ldots,2n$, are complex constants and the indices are elements of
$\Z/(2n+1)\Z$, i.e.\ $\ph_{k+2n+1}=\ph_k$.

We remark that setting $n=1$ in \eqref{eq:sP4h} yields the s\PIV\ system
\begin{subequations}\label{eq:sA2}\begin{align}
\deriv{\ph_0}{z} &+ \ph_0(\ph_1-\ph_2)+2\mu_0=0,\\
\deriv{\ph_1}{z} &+ \ph_1(\ph_2-\ph_0)+2\mu_1=0,\\
\deriv{\ph_2}{z} &+ \ph_2(\ph_0-\ph_1)+2\mu_2=0, 
\end{align} with constraints \begin{equation} 
\ph_0+\ph_1+\ph_2=-2z, \qquad \mu_0+\mu_1+\mu_2=1.
\end{equation}\end{subequations}
Eliminating 
$\ph_1$ and $\ph_2$ in (\ref{eq:sA2}) and setting $\ph_0=w$ yields
\PIV\ (\ref{eq:PIV}) with $\a=\mu_1-\mu_2$ and $\b=-2\mu_0^2$. The s\PIV\ system
(\ref{eq:sA2}) was known to Bureau \citep{refBureau80},
it was subsequently discovered by Adler \citep{refAdler} in the context of periodic
chains of B\"acklund transformations, see also Veselov and Shabat \citep{refVesSh}. The
s\PIV\ system (\ref{eq:sA2}) has important applications in random matrix theory,
which are discussed by Forrester and Witte \citep{refFW01}.

Filipuk and Clarkson \cite{refFilPAC} studied rational solutions of the s\PIV\ hierarchy (\ref{eq:sP4h}) and expressed them in terms of the  \textit{symmetric Hermite polynomials} and  \textit{symmetric Okamoto polynomials}, which respectively are defined in Definitions \ref{def:symherm} and \ref{def:symok} below and are extensions of the generalized Hermite and Okamoto polynomials.

\begin{definition}{\label{def:symherm}\rm The \textit{symmetric Hermite polynomial}\ $H_{\boldsymbol{k}_{2n}}(z)$ is defined by
\begin{equation}
H_{\boldsymbol{k}_{2n}}(z)=\mathcal{W}\left(\big\{H_{k_1+j}(z)\big\}_{j=0}^{k_2-1},\big\{H_{k_1+k_2+k_3+j}(z)\big\}_{j=0}^{k_4-1},\ldots,
\big\{H_{k_1+k_2+\ldots+k_{2n-1}+j}(z)\big\}_{j=0}^{k_{2n}-1}\right),\end{equation}
where $\boldsymbol{k}_{2n}=(k_1,k_2,\ldots,k_{2n})$ and $H_{k}(z)$ is the $k$th Hermite polynomial,
and has degree
$$\mbox{deg}(H_{\boldsymbol{k}_{2n}}(z))=k_{2n}k_{2n-1}+(k_{2n}+k_{2n-2})k_{2n-3}+\ldots + (k_{2n}+k_{2n-2}+\ldots+k_2)k_1.$$
}\end{definition}

\begin{definition}{\label{def:symok}\rm The \textit{symmetric Okamoto polynomial}\ $\GG_{\boldsymbol{k}_{n}}(z)$ is defined by
\begin{equation}
\GG_{\boldsymbol{k}_{n}}(z)=\mathcal{W}\left(\big\{H_{(n+1)j+1}(z)\big\}_{j=0}^{k_1-1},\big\{H_{(n+1)j+2}(z)\big\}_{j=0}^{k_2-1},\ldots,\big\{H_{(n+1)j+n}(z)\big\}_{j=0}^{k_{n}-1}\right),
\end{equation}
where $\boldsymbol{k}_{n}=(k_1,k_2,\ldots,k_{n})$ and $H_{k}(z)$ is the $k$th Hermite polynomial,
and has degree
$$\mbox{deg}(\GG_{\boldsymbol{k}_{n}}(z))=\tfrac12n\sum_{j=1}^nk_j(k_j+1)-\sum_{j=2}^n\sum_{i=1}^{j-1}k_ik_j+\sum_{j=2}^n(1-j)k_j.$$
}\end{definition}

We note that $H_{\boldsymbol{k}_{2}}(z)=H_{k_1,k_2}(z)$, the generalized Hermite polynomial, and $\GG_{\boldsymbol{k}_{2}}(z)=\GG_{k_1,k_2}(z)$, the generalized Okamoto polynomial. Filipuk and Clarkson \cite{refFilPAC} show that pairs of symmetric Hermite polynomials and pairs of symmetric Okamoto polynomials satisfy bilinear equations similar to the forms (\ref{eq:bilinHmn}) and (\ref{eq:bilinQmn}), respectively.

\section{\label{sec:vortex}Point Vortex Dynamics}
The equations of motion for $N$ point vortices with circulations $\G{j}$ at positions $z_j$, $j=1,2,\ldots,N$, 
with no external flow field, are
\begin{equation}
\deriv{\zz_j}{t}=\frac1{2\pi\i} \ssum_{k=1}^{N}\frac{\G{k}}{z_j-z_k},\qquad j=1,2,\ldots,N.\label{eq31}
\end{equation}

\subsection{Identical vortices on a line}
If a vortex configuration rotates as a rigid body with angular velocity $\Omega$ then
\begin{equation*}\deriv{\zz_j}{t} =-\i\Omega \zz_j,\qquad j=1,2,\ldots,N,\label{eq0}\end{equation*}
and so from (\ref{eq31})
\begin{equation*}\la\zz_j = \ssum_{k=1}^{N}\frac{\G{k}}{z_j-z_k},\qquad
j=1,2,\ldots,N, \label{eq33}\end{equation*}
where $\la=2\pi\Omega$. 
Suppose that $z_j$ are real, so $z_j=\zz_j=x_j$, and all the $\G{j}$ are equal, so set $\G{j}=1$ for $j=1,2,\ldots,N$, without loss of generality (by rescaling $x_j$, if necessary), and so we obtain
\begin{equation} x_j=\ssum_{k=1}^{N}\frac1{x_j-x_k},\qquad
j=1,2,\ldots,N,\label{eq2}\end{equation}
which are known as \textit{Stieltjes relations} \citep{refStieltjes85a,refStieltjes85b}. Further, as shown by Stieltjes \citep{refStieltjes85a,refStieltjes85b}, the solutions ${x_1,x_2,\ldots,x_N}$ of equations (\ref{eq2}) are the roots of the $N$th Hermite polynomial $H_N(x)$. This highlights the connection between point patterns corresponding to a relative equilibrium and functions from classical mathematical physics.

\subsection{\label{ssec:3.2}Vortices of the same strength and mixed signs}
Now we consider the situation when the vortices at $z_1,z_2,\ldots,z_{m}$ have strength $\Gamma>0$, i.e.\ there are $m$ positive vortices, and the vortices at $z_{m+1},z_{m+2},\ldots,z_{m+n}$ have strength $-\Gamma$, i.e.\ there are $n$ negative vortices, where $m+n=N$, so that
\begin{equation}\label{eq:gammak} \G{k}=\begin{cases}\Gamma,&\quad \mbox{for}\quad k=1,2,\ldots,m\\
-\Gamma,&\quad \mbox{for}\quad k=m+1,m+2,\ldots,m+n.\end{cases}\end{equation}
We shall consider two important cases of equation (\ref{eq31}) with these circulations.

\begin{enumerate}\item \textit{Stationary vortex patterns} that arise when $\ds\ideriv{\zz_j}{t}=0$, which we discuss in \S\ref{sec:svp}.
\item \textit{Translating vortex patterns} that arise when $\ds\ideriv{\zz_j}{t}=v^*$, with $v^*$ a (complex) constant, which we discuss in \S\ref{sec:tvp}.
\end{enumerate}

\noindent
To study these vortex patterns we define the polynomials
\begin{equation} P(z)=\prod_{j=1}^{m}(z-z_j),\qquad  Q(z)=\prod_{k=1}^{n}(z-\z_{k}),\label{eqPQ}\end{equation}
with $\z_k=z_{k+m}$ for $k=1,2,\ldots,n$, which respectively have roots at ${z_1,z_2,\ldots,z_m}$, i.e.\ the locations of the positive vortices, and ${\z_1,\z_2,\ldots,\z_n}$, i.e.\ the locations of the negative vortices. 

\def\ra{\quad\Rightarrow\quad}
\subsubsection{\label{sec:svp}Stationary vortex patterns}
In this case, setting $\ds\ideriv{\zz_j}{t}=0$  in (\ref{eq31}) gives
$$  \ssum_{k=1}^{m+n}\frac{\G{k}}{z_j-z_k}=0,\qquad j=1,2,\ldots,m+n,$$  with $\G{k}$ given by (\ref{eq:gammak}).
If $P(z)$ and $Q(z)$ are given by (\ref{eqPQ}) then it can be shown that
they satisfy the bilinear equation 
\begin{equation}\deriv[2]{P}{z}Q-2\deriv{P}{z}\deriv{Q}{z}+P\deriv[2]{Q}{z}=0\label{eq:svp1}\end{equation}
(for details see \cite{refAref07a, refANSTV,refKC87}).
As noted in \S\ref{ssec:AM}, polynomial solutions of equation (\ref{eq:svp1}) were studied by Burchnall and Chaundy \citep{refBC29}, and later by Adler and Moser \citep{refAdMoser} --- recall equation (\ref{eq26}), see also Theorem \ref{thm:am1}. In the vortex dynamics literature equation (\ref{eq:svp1}) is sometimes known as Tkachenko's equation (cf.\ \citep{refAref07a,refONeil07b}).
Substituting $P(z)=z^{j}+a_{j-1}z^{j-1}+\ldots +a_0$ and $Q(z)=z^{k}+b_{k-1}z^{k-1}+\ldots +b_0$ into equation  (\ref{eq:svp1}) yields the leading order term
$$[j(j-1)-2jk+k(k-1)]z^{j+k-2}=[(j-k)^2-(j+k)]z^{j+k-2},$$ thus $j-k=n$ and $j+k=n^2$ and so
$j=\tfrac12n(n+1)$ and $k=\tfrac12n(n-1)$, with $n\in\Z$. Thus the degrees of the polynomials $P(z)$ and $Q(z)$ are two successive triangular numbers.

Burchnall and Chaundy \citep{refBC29} and Adler and Moser \citep{refAdMoser} showed that polynomial solutions of equation (\ref{eq:svp1}) are expressed in terms of the \textit{\AM\ polynomials}, which were defined in \S\ref{ssec:AM}, as given in the following result; see also \citep{refANSTV,refBartman}.

\begin{theorem}{\label{thm:am1}Polynomial solutions of equation (\ref{eq:svp1}) are given by
$P(z)=\Th_{n+1}(z;\bfk_{n+1})$ and $Q(z)=\Th_{n}(z;\bfk_{n})$,
where  $\Th_{n}(z;\bfk_{n})$ is the \AM\  polynomial. }\end{theorem}

Adler and Moser \citep{refAdMoser}, using ideas on Darboux transformations due to Crum \citep{refCrum}, 
also proved the following result, which led to the Wronskian representation of the \AM\ polynomials given in Definition \ref{def:ampoly}.

\begin{theorem}{The \sch\ equation
\begin{equation}-\deriv[2]{\psi}{z}+u(z)\psi=0,\label{eq:svp2}\end{equation} with potential $\ds u(z)=-2\deriv[2]{}{z}\ln\Th_n(z;\bfk_{n})$, where  $\Th_{n}(z;\bfk_{n})$ is the \AM\ polynomial,
has the eigenstate 
\begin{equation}\psi(z)=\ifrac{\Th_{n+1}(z;\bfk_{n+1})}{\Th_n(z;\bfk_{n})}.\end{equation}
}\end{theorem}

We remark that substituting $$ u(z)=-2\deriv[2]{}{z}\ln Q(z),\qquad\psi(z)=\frac{P(z)}{Q(z)},$$ into the \sch\ equation (\ref{eq:svp2}) yields the bilinear equation (\ref{eq:svp1}).

\subsubsection{\label{sec:tvp}Translating vortex patterns}
In this case, setting $\ds\ideriv{\zz_j}{t}=v^*$, with $v^*$ a constant, in (\ref{eq31}) yields
$$  \frac1{2\pi\i}\ssum_{k=1}^{m+n}\frac{\G{k}}{z_j-z_k}=v^*,\qquad j=1,2,\ldots,m+n,$$ with $\G{k}$ given by (\ref{eq:gammak}). 
If $P(z)$ and $Q(z)$ are given by (\ref{eqPQ}) then it can be shown that
they satisfy the bilinear equation
\begin{equation}\deriv[2]{P}{z}Q-2\deriv{P}{z}\deriv{Q}{z}+P\deriv[2]{Q}{z}+2\mu\left(\deriv{P}{z}Q-P\deriv{Q}{z}\right)=0,\label{eq:tvp1}\end{equation}
where $\mu=2\pi\i v^*/\Gamma$\ (for details see \cite{refAref07a, refANSTV,refKC87}).
Substituting $P(z)=z^{j}+a_{j-1}z^{j-1}+\ldots +a_0$ and $Q(z)=z^{k}+b_{k-1}z^{k-1}+\ldots +b_0$ into equation  (\ref{eq:tvp1}) yields the leading order term $2\mu(j-k)z^{j+k-1}$ and so $j=k$, i.e.\ the polynomials $P(z)$ and $Q(z)$ have the same degree. Theorem \ref{thm:3.3} shows that the degrees of the polynomials $P(z)$ and $Q(z)$ are triangular numbers, though it appears not to be understood why this restriction holds, independent of knowing the solution, in contrast to the situation for stationary vortex patterns in \S\ref{sec:svp}.

Adler and Moser \citep{refAdMoser} proved the following result concerning polynomial solutions of equation (\ref{eq:tvp1}); see also \citep{refANSTV,refBartman,refLout04}.

\begin{theorem}{\label{thm:3.3}Polynomial solutions of equation (\ref{eq:tvp1}) are given by
$P(z)=\Tht_{n}(z;\mu,\bfk_{n})$ and $Q(z)=\Th_{n}(z;\bfk_{n})$,
where $\Th_{n}(z;\bfk_{n})$ is the \AM\  polynomial and
$\Tht_{n}(z;\mu,\bfk_{n})$ is the modified \AM\ polynomial.
Further \sch\ equation
\begin{equation}-\deriv[2]{\psi}{z}+u(z)\psi=-\mu^2\psi,\label{tvp2}\end{equation} with potential $\ds u(z)=-2\deriv[2]{}{z}\ln\Th_n(z;\bfk_{n})$, where $\Th_n(z;{\bfk}_{n})$ is the \AM\ polynomial,  has the eigenstate \begin{equation}\psi(z)=\ifrac{\Tht_{n}(z;\mu,{\bfk}_{n})}{\Th_n(z;\bfk_{n})},\end{equation}
where $\Tht_{n}(z;\mu,\bfk_{n})$ is the modified \AM\ polynomial. }\end{theorem}

We remark that substituting $$u(z)=-2\deriv[2]{}{z}\ln Q(z),\qquad \psi(z)=\frac{P(z)\,\e^{\mu z}}{Q(z)},$$ into the \sch\ equation (\ref{tvp2}) 
yields equation (\ref{eq:tvp1}).

\subsection{\label{ssec:3.3}Quadrupole Background Flow}
The equations of motion for $N$ point vortices with circulations $\G{j}$ at positions $z_j$, $j=1,2,\ldots,N$ in a background flow $w(z)$ are
\begin{equation}\deriv{\zz_j}{t}=\frac1{2\pi\i} \ssum_{k=1}^{N}\frac{\G{k}}{z_j-z_k}+ \frac{w^*(z_j)}{2\pi\i},\qquad j=1,2,\ldots,N.\end{equation}
Now suppose that $\ds\ideriv{\zz_j}{t}=0$, i.e.\ stationary flow, $w(z)=v^*z^*$, with $v^*$ a (complex) constant, i.e.\ quadrupole background flow, and $\G{k}$ is given by (\ref{eq:gammak}). 
In this case, setting $v=\mu\Gamma$ gives 
\begin{equation} \ssum_{k=1}^{m+n}\frac{\G{k}}{z_j-z_k}=-\Gamma\mu z_j,\qquad j=1,2,\ldots,m+n,\end{equation}  and so using (\ref{eq:gammak}) yields
\begin{subequations}\label{eq:dagger}
\begin{align}&\ssum_{k=1}^{m} \frac1{z_j-z_k}=-\mu z_j+\sum_{\ell=1}^{n}\frac1{z_j-\z_\ell},&& j=1,2,\ldots,m,\label{eq;dag}\\
&\ssum_{\ell=1}^{n} \frac1{\z_j-\z_k}=\mu\z_j+\sum_{k=1}^{m}\frac1{\z_ j-z_k},&& j =1,2,\ldots,n.
\label{eq;dagg}\end{align}\end{subequations}
If $P(z)$ and $Q(z)$ are given by (\ref{eqPQ}) then 
\begin{align*}
\deriv{P}{z}&= P \sum_{j=1}^{m}\frac1{z-z_j},\qquad \deriv{Q}{z}= Q \sum_{k=1}^{n}\frac1{z-\z_k}, \\
\deriv[2]{P}{z}&= 2P\sum_{j=1}^{m}\ssum_{k=1}^{m} \frac1{z-z_j}\frac1{z_j-z_k}
= 2P\sum_{j=1}^{m} \frac1{z-z_j}\left(-\mu z_j+\sum_{k=1}^{n}\frac1{z_j-\z_k}\right),\\
\deriv[2]{Q}{z}&= 2Q\sum_{k=1}^{n}\ssum_{\ell=1}^{n} \frac1{z-\z_k}\frac1{\z_k-\z_\ell} 
= 2Q\sum_{k=1}^{n} \frac1{z-\z_k}\left(\mu\z_k+\sum_{\ell=1}^{m}\frac1{\z_k-z_\ell}\right),
\end{align*}
using (\ref{eq:dagger}) and so
\[\begin{split}
Q\deriv[2]{P}{z}+P\deriv[2]{Q}{z} &=2PQ\sum_{j=1}^{m}\sum_{k=1}^{n} \frac1{z_j-\z_k}\left(\frac1{z-z_j} -\frac1{z-\z_k}\right) -2\mu PQ\left(\sum_{j=1}^{m}\frac{z_j}{z-z_j}-\sum_{k=1}^{n}\frac{\z_k}{z-\z_k}\right)\\ &=2PQ\sum_{j=1}^{m}\frac1{z-z_j}\sum_{k=1}^{n}\frac1{z-\z_k} -2\mu PQ\left\{\sum_{j=1}^{m}\frac{z-(z-z_j)}{z-z_j}-\sum_{k=1}^{n}\frac{z-(z-\z_k)}{z-\z_k}\right\}\\
&=2\deriv{P}{z}\deriv{Q}{z}-2\mu PQ\left\{z\sum_{j=1}^{m}\frac{1}{z-z_j}-z\sum_{k=1}^{n}\frac{1}{z-\z_k}+(m-n)\right\}\\
&=2\deriv{P}{z}\deriv{Q}{z}-2\mu z\left(\deriv{P}{z}Q-P\deriv{Q}{z}\right)+2\mu(m-n)PQ,\end{split}\]
and so the polynomials $P(z)$ and $Q(z)$ satisfy the bilinear equation
\begin{equation}\label{eq:qbf1}\deriv[2]{P}{z}Q-2\deriv{P}{z}\deriv{Q}{z}+P\deriv[2]{Q}{z}+2\mu z\left(\deriv{P}{z}Q-P\deriv{Q}{z}\right)-2\mu(m-n) PQ=0.\end{equation}
Substituting $P(z)=z^{m}+a_{m-1}z^{m-1}+\ldots +a_0$ and $Q(z)=z^{n}+b_{n-1}z^{n-1}+\ldots +b_0$ into equation (\ref{eq:qbf1}) shows that the leading order term is identically satisfied 
and so the degrees of the polynomials $P(z)$ and $Q(z)$ are not restricted.

Kadtke and Campbell \citep{refKC87} obtained some polynomial solutions of equation (\ref{eq:qbf1}) in the case when $m=n=2$ and $m=n=4$, though they also claimed that there were no solutions when $m=n=6$. However, we have found 6 pairs of solutions of equation (\ref{eq:qbf1}) in the case when $m=n=6$ and 12 pairs of solutions when $m=n=8$ (for this case 2 pairs of solutions are given in \citep{refKC87}). The roots of these polynomials are invariant under reflections in the real and imaginary $z$-axes when $m=n$. In Tables \ref{tab41} and \ref{tab42} some polynomial solutions of equation (\ref{eq:qbf1}) for $\mu=-\tfrac12$ in the cases when $m=n$ and $m=n+1$ are given, respectively. All of these polynomials are Wronskians of the scaled Hermite polynomial, in fact they have the form (\ref{sol:qbf11}), with $H_{k_j}(z)$ replaced by $H_{k_j}(z/\sqrt2)$ and scaled so that the polynomials are monic.

\begin{table}[!htp]
$$ \begin{array}{|l@{\quad}|@{\quad}l@{\quad}|@{\quad}l|}\hline
& P(z) & Q(z) \\ \hline
m=n=2 &  z^2+1 &  z^2-1\\ \hline
m=n=3 &  z^3+3z &  z^3\\
  & z^3 &  z^3-3z \\ \hline
m=n=4 &  z^4+6z^2+3 &  z^4+2z^2-1\\
 &  z^4+2z^2-1 &  z^4-2z^2-1\\
  & z^4-2z^2-1 &  z^4-6z^2+3 \\ \hline
m=n=5 &  z^5+10z^3+15z &  z^5+5z^3\\
  & z^5+5z^3 &  z^5-5z \\ 
  &  z^5-5z & z^5-5z^3  \\ 
  & z^5-5z^3  & z^5-10z^3+15z \\ 
 & z^5+2z^3+3z & z^5-2z^3+3z \\   \hline
m=n=6 & z^6+15z^4+45z^2+15 & z^6+9z^4+9z^2-3\\
  & z^6+9z^4+9z^2-3 & z^6+3z^4-9z^2-3\\
  & z^6+3z^4-9z^2-3 & z^6-3z^4-9z^2+3\\
  & z^6-3z^4-9z^2+3 & z^6-9z^4+9z^2+3\\
  & z^6-9z^4+9z^2+3 & z^6-15z^4+45z^2-15\\
  & z^6+3z^4+9z^2-9 & z^6-3z^4+9z^2+9\\   \hline
m=n=7 &  z^7+21z^5+105z^3+105z &  z^7+14z^5+35z^3\\
  & z^7+14z^5+35z^3 & z^7+7z^5-7z^3-21z \\ 
  &  z^7+7z^5-7z^3-21z & z^7-21z^3  \\ 
  & z^7-21z^3  & z^7-7z^5-7z^3+21z \\ 
 & z^7-7z^5-7z^3+21z & z^7-14z^5+35z^3 \\  
 & z^7-14z^5+35z^3 & z^7-21z^5+105z^3-105z \\  
 &z^7+9z^5+15z^3+15z & z^7+3z^5-3z^3+3z \\
 & z^7+6z^5+15z^3 & z^7-z^5+5z^3+15z  \\
 & z^7+z^5+5z^3-15z & z^7-6z^5+15z^3 \\ \hline
m=n=8  &  z^{8}+28z^{6}+210z^{4}+420z^{2}+105 &  z^{8}+20z^{6}+90z^{4}+60z^{2}-15 \\
 &  z^{8}+20z^{6}+90z^{4}+60z^{2}-15 &  z^{8}+12z^{6}+10z^{4}-60z^{2}-15\\
 &  z^{8}+12z^{6}+10z^{4}-60z^{2}-15  &  z^{8}+4z^{6}-30z^{4}-36z^{2}+9\\
 &  z^{8}+4z^{6}-30z^{4}-36z^{2}+9 &  z^{8}-4z^{6}-30z^{4}+36z^{2}+9 \\
 &  z^{8}-4z^{6}-30z^{4}+36z^{2}+9 &  z^{8}-12z^{6}+10z^{4}+60z^{2}-15 \\
 &  z^{8}-12z^{6}+10z^{4}+60z^{2}-15 &  z^{8}-20z^{6}+90z^{4}-60z^{2}-15 \\
 &  z^{8}-20z^{6}+90z^{4}-60z^{2}-15 &  z^{8}-28z^{6}+210z^{4}-420z^{2}+105 \\
 &  z^{8}+10z^{6}+30z^{4}+30z^{2}-15 &  z^{8}+2z^{6}+30z^{2}+15 \\
 &  z^{8}+4z^{6}+6z^{4}-12z^{2}-3 &  z^{8}-4z^{6}+6z^{4}+12z^{2}-3 \\
 &  z^{8}-2z^{6}-30z^{2}+15 &  z^{8}-10z^{6}+30z^{4}-30z^{2}-15 \\
 &  z^{8}+8z^{6}+30z^{4}+45 &  z^{8}+10z^{4}-15\\
&  z^{8}+10z^{4}-15 &  z^{8}-8z^{6}+30z^{4}+45 \\
   \hline
\end{array}$$\caption{\label{tab41}Some polynomial solutions of equation (\ref{eq:qbf1}) for $\mu=-\tfrac12$ and $m=n$}\end{table}

\begin{table}[!htp]
$$\begin{array}{|c@{\quad}|@{\quad}l@{\quad}|@{\quad}l|}\hline
& P(z) & Q(z) \\ \hline 
m=3, n=2 &  z^3+3z & z^2+1\\ \hline 
m=4,n=3 &  z^4+3 &  z^3-3z\\
 &  z^4+6z^2+3 &  z^3+3z\\ \hline 
m=5,n=4 &  z^5+10z^3+15z &  z^4+6z^2+3\\
& z^5-2 z^3+3 z & z^4-6 z^2+3\\
& z^5+2 z^3+3 z & z^4-2 z^2-1\\ \hline 
m=6,n=5 &  z^6+3z^4+9z^2-9  &  z^5-2z^3+3z\\
&  z^6+5z^4+5z^2+5 & z^5-5z\\
&  z^6+15z^4+45z^2+15 &  z^5+10z^3+15z\\
&  z^6-5z^4+5z^2-5&  z^5-10z^3+15z\\
&  z^6 &  z^5-5z^3\\ \hline 
m=7,n=6 &   z^7+3z^5-3z^3+3z &  z^6-3z^4-9z^2+3\\
 &  z^7+9z^5+15z^3+15z&  z^6+3z^4-9z^2-3\\
&  z^7-3z^5-3z^3-3z&  z^6-9z^4+9z^2+3\\
&  z^7-9z^5+15z^3-15z & z^6-15z^4+45z^2-15 \\
&  z^7+21z^5+105z^3+105z&   z^6+15z^4+45z^2+15\\
&  z^7+z^5+5z^3-15z &  z^6-5z^4+5z^2-5\\ \hline
m=8,n=7 & z^{8}+28z^{6}+210z^{4}+420z^{2}+105 & z^7+21z^5+105z^3+105z \\
& z^{8}+14z^{6}+42z^{4}+42z^{2}+21 & z^7+7z^5-7z^3-21z  \\
& z^{8}+4z^{6}+6z^{4}-12z^{2}-3 & z^7-3z^5-3z^3-3z \\
& z^{8}+10z^{6}+30z^{4}+30z^{2}-15 & z^7+3z^5-3z^3+3z \\
& z^{8}-14z^{6}+42z^{4}-42z^2+21 & z^7-21z^5+105z^3-105z \\
& z^{8}+8z^{6}+30z^{4}+45 & z^7+z^5+5z^3-15z \\
& z^{8}-2z^{6}-30z^{2}+15  & z^7-9z^5+15z^3-15z \\
& z^{8}-14z^{4}-7 & z^7-7z^5-7z^3+21z \\
& z^{8}+7z^{6} & z^7-21z^3 \\
& z^{8}-7z^6 & z^7-14z^5+35z^3 \\
\hline \end{array}$$
\caption{\label{tab42}Some polynomial solutions of equation (\ref{eq:qbf1}) for $\mu=-\tfrac12$ and $m=n+1$}
\end{table}

We note that equation (\ref{eq:qbf1}) for $\mu=-1$\ (which we can assume without loss of generality, by rescaling $z$ if necessary), i.e.\
\begin{equation}\label{eq:qbf11}
\deriv[2]{P}{z}Q-2\deriv{P}{z}\deriv{Q}{z}+P\deriv[2]{Q}{z}-2z\left(\deriv{P}{z}Q-P\deriv{Q}{z}\right)+2 (m-n)PQ=0,
\end{equation}
can be written in the form
\begin{equation}\label{eq:qbf12}
\left[\D_z^2-2z\D_z+2(m-n)\right]P\cdot Q=0,
\end{equation}with $\D_z$ the Hirota operator (\ref{Hirop}). Hence from (\ref{eq:bilinHmn}) and (\ref{eq:bilinQmn}), we can obtain three sets of solutions of equation (\ref{eq:qbf11}), or equivalently (\ref{eq:qbf12}), in terms of the generalized Hermite polynomials $H_{j,k}(z)$ and the generalized Okamoto polynomials $\GG_{j,k}(z)$, respectively defined in Definitions \ref{genherm} and \ref{genokpolytilde}, which are given in Table \ref{tab:genHermgenOk}. As shown below, it turns out that only two of these sets have the form (\ref{sol:qbf11}). Further, Filipuk and Clarkson \cite{refFilPAC} show that there are solutions of equation (\ref{eq:qbf11}), or equivalently (\ref{eq:qbf12}), in terms of the symmetric Hermite polynomials $H_{\boldsymbol{k}_{n}}(z)$ and the symmetric Okamoto polynomials $\GG_{\boldsymbol{k}_{n}}(z)$, which are defined in Definitions \ref{def:symherm} and \ref{def:symok}, respectively.

 \begin{table}[!htp]
\[\begin{array}{|@{\ }c@{\ }|@{\ }c@{\ }|@{\ }c@{\ }|}\hline
P(z) & Q(z) & m - n \\ \hline
H_{j,k}(z) & H_{j+1,k}(z)   & -k  \\
H_{j,k+1}(z) & H_{j,k}(z)  & j \\
H_{j,k+1}(z) & H_{j+1,k}(z)  & j-k \\ \hline 
\GG_{j+1,k}(z) & \GG_{j,k}(z) & j-2k+2  \\
\GG_{j,k+1}(z) & \GG_{j,k}(z) & -2j+k+1  \\
\GG_{j,k}(z) & \GG_{j+1,k+1}(z) &-j-k-2 \\ \hline 
\end{array}\]
\caption{\label{tab:genHermgenOk}Solutions of (\ref{eq:qbf11}) and (\ref{eq:qbf12}) in terms of generalized Hermite and generalized Okamoto polynomials.}\end{table}

\def\N{\ell}
\subsubsection{Relationship to the \sch\ equation}
In \S\ref{ssec:3.2} it was shown that stationary and translating vortex equilibrium solutions can be expressed in terms of the roots of the \AM\ and modified \AM\ polynomials and these were related to eigenstates of the \sch\ equations (\ref{eq:svp2}) and (\ref{tvp2}). Now in order to describe quadrupole background flow equilibrium solutions we shall relate solutions of equation (\ref{eq:qbf11}) to eigenstates of a \sch\ equation.

Consider the \sch\ operator
\begin{equation}\label{sch:op}
\mathcal{L} = -\deriv[2]{}{z}+u(z),\end{equation}
with a potential $u(z)$ which is meromorphic in $\Com$. The \sch\ operator $\mathcal{L}$ (\ref{sch:op}) is said to have \textit{trivial monodromy} if all the solutions of the corresponding  \sch\ equation
\begin{equation}\mathcal{L} \psi=-\deriv[2]{\psi}{z}+u(z)\psi=\la\psi,\label{eq:sch}\end{equation}
are also meromorphic in $\Com$ for \textit{all}\ $\la$. Further, such an operator is said to be \textit{monodromy-free}.

Duistermaat and Gr\"unbaum \citep{refDG86} considered the problem of the classification of \sch\ operators for rational potentials decaying at infinity, i.e.\ (\ref{sch:op}) with $\lim_{|z|\to\infty}u(z)=0$, and obtained the following result.

\begin{theorem}{\label{thm:DG}Every \sch\ operator $\mathcal{L}$ with trivial monodromy for rational potentials decaying at infinity has the form
\begin{equation}\label{eq:DG} \mathcal{L} = -\deriv[2]{}{z}+\sum_{j=1}^N\frac{\nu_j(\nu_j+1)}{(z-z_j)^2},\end{equation}
where $\nu_1,\nu_2,\ldots,\nu_N$ are integers. Further all such potentials are obtained by finitely many Darboux transformations applied to the zero potential.}\end{theorem}

Oblomkov \citep{refOblomkov} generalized Theorem \ref{thm:DG} to the case of rational potentials with quadratic growth at infinity and obtained the following result. 

\begin{theorem}{\label{thm:Ob}Every \sch\ operator $\mathcal{L}$ with trivial monodromy, and with a quadratically increasing rational potential, 
i.e.\ $u(z)=z^2+R(z)$, with $\lim_{|z|\to\infty}R(z)=0$,
has the form
\begin{equation} \mathcal{L} = -\deriv[2]{}{z}+z^2-2\deriv[2]{}{z}\ln\mathcal{W}\left(H_{k_1}(z),H_{k_2}(z),\ldots, H_{k_\N}(z)\right),\end{equation}
where $k_1,k_2,\ldots,k_\N$ are a sequence of distinct positive integers and $H_{k}(z)$ is the $k$th Hermite polynomial. Further all such potentials are obtained obtained by finitely many Darboux transformations applied to the potential $u=z^2$.
}\end{theorem}

Then, using Theorem \ref{thm:Ob} and a corollary of results on Darboux transformations due to Crum \citep{refCrum} we have the following result (see also Veselov \citep{refVeselov}).

\begin{theorem}{The \sch\ equation (\ref{eq:sch}) with potential
\begin{equation}\label{sch:pot}
u(z)=z^2-2\deriv[2]{}{z}\ln\mathcal{W}\left(H_{k_1}(z),H_{k_2}(z),\ldots,H_{k_\N}(z)\right),
\end{equation}
where $k_1,k_2,\ldots,k_\N$ are a sequence of distinct positive integers and $H_{k}(z)$ is the $k$th Hermite polynomial, has the eigenstates 
\begin{subequations}\begin{align}
\psi(z)&=\frac{\mathcal{W}\left(H_{k_1}(z),H_{k_2}(z),\ldots,H_{k_\N}(z),H_{k_{\N+1}}(z)\right)}{\mathcal{W}\left(H_{k_1}(z),H_{k_2}(z),\ldots,H_{k_\N}(z)\right)}\exp\left(-\tfrac12z^2\right),\\
\psi(z)&=\frac{\mathcal{W}\left(H_{k_1}(z),H_{k_2}(z),\ldots,H_{k_{\N-1}}(z)\right)}{\mathcal{W}\left(H_{k_1}(z),H_{k_2}(z),\ldots,H_{k_\N}(z)\right)}\exp\left(\tfrac12z^2\right),
\end{align}\end{subequations}
with $k_{\N+1}$ a positive integer such that $k_{\N+1}\not=k_j$, for $j=1,2,\ldots,\N$, 
 for the eigenvalues $\la=1+2(k_{\N+1}-\N)$ and $\la=2(\N-k_{\N-1})-1$, respectively.}
\end{theorem}

It should be noted that the potential (\ref{sch:pot}) may possess singularities on the real axis, so that the terms ``eigenstate'" and ``eigenvalue" should not be understood in quantum mechanics sense. The associated regularity conditions have been studied by Adler \citep{refAdler2} and Dubov, Eleonskii, and Kulagin \citep{refDEK}.

We remark that substituting $$ u(z)=z^2-2\deriv[2]{}{z}\ln Q(z),\qquad \psi(z)=\frac{P(z)}{Q(z)}\exp(-\tfrac12z^2),$$ into the \sch\ equation (\ref{eq:sch}) yields equation (\ref{eq:qbf11}) with $m-n=\lambda-1$. Consequently we have the following result (see also Loutsenko \citep[Proposition 4]{refLout03}).

\begin{theorem}{\label{thm:3.7}The bilinear equation (\ref{eq:qbf11})
has polynomial solutions in the form
\begin{subequations}\label{sol:qbf11}\begin{align}
P(z) &= \mathcal{W}\left(H_{k_1}(z),H_{k_2}(z),\ldots,H_{k_\N}(z),H_{k_{\N+1}}(z)\right),\\
Q(z) &= \mathcal{W}\left(H_{k_1}(z),H_{k_2}(z),\ldots,H_{k_\N}(z)\right),
\end{align}\end{subequations}
where $k_1,k_2,\ldots,k_\N,k_{\N+1}$ are a sequence of {distinct} positive integers and $H_{k}(z)$ is the $k$th Hermite polynomial. Further the degrees of the polynomials $P(z)$ and $Q(z)$, respectively $m$ and $n$, are given by
\[m= \sum_{j=1}^{\N+1} k_j - \tfrac12\N(\N+1),\qquad
n=\sum_{j=1}^{\N} k_j - \tfrac12\N(\N-1)\]
i.e.\ $m-n = k_{\N+1}-\N$.}\end{theorem}

We note that the polynomials $P(z)$ and $Q(z)$ given by (\ref{sol:qbf11}) do not involve arbitrary parameters, in contrast to the \AM\ polynomials which give polynomial solutions of equations (\ref{eq:svp1}) and (\ref{eq:tvp1}).

Theorem \ref{thm:3.7} would appear to give a complete classification of polynomial solutions of the bilinear equation (\ref{eq:qbf11}). As remarked above, all the polynomials in Tables \ref{tab41} and \ref{tab42} have the form (\ref{sol:qbf11}), with $H_{k_j}(z)$ replaced by $H_{k_j}(z/\sqrt2)$ and scaled so that the polynomials are monic.

However there are additional polynomial solutions of equation (\ref{eq:qbf11}) in terms of the generalized Hermite polynomials $H_{k_1,k_2}(z)$ and the generalized Okamoto polynomials $\GG_{k_1,k_2}(z)$. In Table \ref{tab:genHermgenOk}, there are three pairs of generalized Hermite polynomials and three pairs of generalized Okamoto polynomials which satisfy equation (\ref{eq:qbf11}), though only two pairs of each are of the form (\ref{sol:qbf11}) in which the wronskian defining $P(z)$ has one more Hermite polynomial than the wronskian defining $Q(z)$. Specifically the polynomials
\begin{subequations}\label{sol:qbf12}
\begin{align}P(z)&= H_{k_1,k_2}(z)\equiv \mathcal{W}\left(H_{k_1},H_{k_1+1},\ldots,H_{k_1+k_2-1}\right),\\ 
Q(z)&= H_{k_1+1,k_2}(z)\equiv \mathcal{W}\left(H_{k_1+1},H_{k_1+2},\ldots,H_{k_1+k_2}\right),
\end{align}\end{subequations}
where the wronskians defining $P(z)$ and $Q(z)$ have the same number of Hermite polynomials, and
\begin{subequations}\label{sol:qbf13}
\begin{align}P(z)&=\GG_{k_1,k_2}(z)\equiv\mathcal{W}(H_{1},H_{4},\ldots,H_{3k_1-2},H_{2},H_{5},\ldots,H_{3k_2-1}),\\ Q(z)&= \GG_{k_1+1,k_2+1}(z)\equiv\mathcal{W}(H_{1},H_{4},\ldots,H_{3k_1+1},H_{2},H_{5},\ldots,H_{3k_2+2}),\end{align}\end{subequations} 
where the wronskian defining $P(z)$ has two fewer Hermite polynomials than the wronskian defining $Q(z)$,
are pairs of solutions of equation (\ref{eq:qbf11}), yet are clearly not of the form (\ref{sol:qbf11}). An analogous result also holds for the symmetric Hermite polynomials $H_{\boldsymbol{k}_{2n}}(z)$ and the symmetric Okamoto polynomials  $\GG_{\boldsymbol{k}_{n}}(z)$. Specifically the polynomials
\begin{subequations}\label{sol:qbf14}
\begin{align}P(z)&=H_{k_1,k_2,\ldots,k_{2n}}(z)\nonumber\\
&\equiv\mathcal{W}\left(\big\{H_{k_1+j}(z)\big\}_{j=0}^{k_2-1},
\big\{H_{k_1+k_2+k_3+j}(z)\big\}_{j=0}^{k_4-1},\ldots,
\big\{H_{k_1+k_2+\ldots+k_{2n-1}+j}(z)\big\}_{j=0}^{k_{2n}-1}\right),\\
Q(z)&=H_{k_1+1,k_2,k_3+1,k_4\ldots,k_{2n-1}+1,k_{2n}}(z)\nonumber\\
& \equiv\mathcal{W}\left(\big\{H_{k_1+j}(z)\big\}_{j=1}^{k_2},\big\{H_{k_1+k_2+k_3+j}(z)\big\}_{j=1}^{k_4},\ldots,
\big\{H_{k_1+k_2+\ldots+k_{2n-1}+j}(z)\big\}_{j=1}^{k_{2n}}\right),\end{align}\end{subequations}
where the wronskians defining $P(z)$ and $Q(z)$ have the same number of Hermite polynomials, and
\begin{subequations}\label{sol:qbf15}
\begin{align}
P(z)&=\GG_{k_1,k_2,\ldots,k_{n}}(z)\nonumber\\
& \equiv\mathcal{W}\left(\big\{H_{(n+1)j+1}(z)\big\}_{j=0}^{k_1-1},\big\{H_{(n+1)j+2}(z)\big\}_{j=0}^{k_2-1},\ldots,
\big\{H_{(n+1)j+n}(z)\big\}_{j=0}^{k_{n}-1}\right),\\
Q(z)&=\GG_{k_1+1,k_2+1,\ldots,k_{n}+1}(z)\nonumber\\ 
&\equiv\mathcal{W}\left(\big\{H_{(n+1)j+1}(z)\big\}_{j=0}^{k_1},\big\{H_{(n+1)j+2}(z)\big\}_{j=0}^{k_2},\ldots,
\big\{H_{(n+1)j+n}(z)\big\}_{j=0}^{k_{n}}\right),
\end{align}\end{subequations}
where the wronskian defining $P(z)$ has $n$ fewer Hermite polynomials than the wronskian defining $Q(z)$,
are pairs of solutions of equation (\ref{eq:qbf11}), yet also are not of the form (\ref{sol:qbf11}). Hence we have demonstrated that there are additional polynomial solutions of equation (\ref{eq:qbf11}) to those given in Theorem \ref{thm:3.7}. The natural question is whether there are any more?

Using MAPLE, we have also found the polynomial solutions of equation (\ref{eq:qbf11}) with $m=n$ and with $m=n+1$, given in Tables \ref{tab:addsols1} and \ref{tab:addsols2}, respectively, which have the property that either the polynomials $P(z)$ or $Q(z)$, or both, have multiple roots, other than $z=0$, and the polynomials $P(z)$ and $Q(z)$ have common roots, other than $z=0$. Further these polynomials do {not} appear to be wronskians of Hermite polynomials.
 \begin{table}[!htp]
\[\begin{array}{|@{\ }c@{\ }|@{\ }c@{\ }|}\hline
P(z) & Q(z) \\ \hline
(z^2+\sqrt{5}\,z+2)^3 & (z^2+\sqrt{5}\,z+2)(z^4+2\sqrt{5}\,z^3+6z^2-3)\\[2.5pt]
\big(z^2+\tfrac34+\tfrac14\i\sqrt{15}\big)^3 & \big(z^2+\tfrac34+\tfrac14\i\sqrt{15}\big)\big[z^4-(\tfrac34-\tfrac14\i\sqrt{15})z^2+\tfrac34\big]\\[2.5pt]
(z+\tfrac12\sqrt{6})^6 & (z^3+\tfrac32\sqrt{6}\,z^2+\tfrac32z-\tfrac74\sqrt{6})(z+\tfrac12\sqrt{6})^3 \\[2.5pt] 
(z^2+1)^3(z^2-1) & (z^2+1)(z^2-1)^3 \\[2.5pt]
z^3(z^2+\tfrac32)^3 & z(z^2+\tfrac32)(z^6-\tfrac32z^4+\tfrac94z^2-\tfrac98) \\[2.5pt]
z\big(z^2+\tfrac54+\tfrac14\i\sqrt{31}\big)^3 & z^3\big(z^2+\tfrac54+\tfrac14\i\sqrt{31}\big)\big(z^2-1+\tfrac12\i\sqrt{31}\big)\\[2.5pt]
(z^2+1)^3(z^2-1) & (z^2+1)(z^2-1)^3 \\[2.5pt]
z^3\big[z^2+\tfrac34(1+\i\sqrt7)\big]^3 & z(z^2-\tfrac32)\big[z^2+\tfrac34(1+\i\sqrt7)\big]
\big[z^4-\tfrac32(1-\i\sqrt7)z^2-\tfrac94\big]\\[2.5pt]
(z^2+\tfrac32)^6 & (z^2+\tfrac32)^3(z^6-\tfrac32z^4+\tfrac{27}{4}z^2-\tfrac98)\\[2.5pt]
(z^2+\tfrac32)^6(z^2-\tfrac32)^3 & (z^2+\tfrac32)^3(z^2-\tfrac32)^6\\[2.5pt]
\hline \end{array}\]
\caption{\label{tab:addsols1}Some additional polynomial solutions of equation (\ref{eq:qbf11}) with $m=n$.}\end{table}
 \begin{table}[!htp]
\[\begin{array}{|@{\ }c@{\ }|@{\ }c@{\ }|}\hline
P(z) & Q(z) \\ \hline
(z+1)^3 & (z+2)(z+1)  \\[2.5pt]
(z^3+2\sqrt3\,z^2+\tfrac92z+\sqrt3)(z+\sqrt3) & (z+\sqrt3)^3\\[2.5pt]
(z^2+\tfrac12)^3 & z(z^2+\tfrac12)(z^2-\tfrac32)\\[2.5pt]
(z^2+\sqrt{10}\,z+3)^3 & (z^2+\sqrt{10}\,z+3)(z^2+\tfrac32\sqrt{10}\,z+6)(z+\tfrac12\sqrt{10})\\[2.5pt]
(z+\tfrac12\sqrt{14})^6 & (z^2+\tfrac32\sqrt{14}\,z+8)(z+\tfrac12\sqrt{14})^3 \\[2.5pt] 
z(z^2-\tfrac32)(z^4+\tfrac{15}4) & (z^2-\tfrac32)^3  \\[2.5pt]
(z+2)^6(z+3) & (z+2)^3(z+3)^3  \\[2.5pt]
(z-2)^6(z-3) & (z-2)^3(z-3)^3  \\[2.5pt]
\hline \end{array}\]
\caption{\label{tab:addsols2}Some additional polynomial solutions of equation (\ref{eq:qbf11}) with $m=n+1$.}\end{table}

\section{\label{sec4}Discussion}
In paper we have investigated the relationship between equilibria of vortex patterns and the roots of polynomials associated with rational solutions of the \p\ and soliton equations. 
The examples which we have considered in this paper illustrate the power of the ``method of polynomials'' when it can be used. The polynomials with roots at the positions of the vortices, are obtained from the basic vortex equilibrium equations and then their roots give the locations of the vortices.

When all the vortex circulations have the same absolute value, stationary vortex configurations and uniformly translating vortex configurations are known to be given by the zeros of the \AM\ polynomials and related to eigenstates of the \sch\ equation for rational potentials decaying at infinity.

For quadrupole background flow and all the vortex circulations have the same absolute value, vortex configurations were shown to be given by the zeros of polynomials expressed as wronskians of Hermite polynomials and related to eigenstates of the \sch\ equation for rational potentials with quadratic growth at infinity. Further using results on the generalized Hermite polynomials $H_{m,n}(z)$ and generalized Okamoto polynomials $\GG_{m,n}(z)$, which are associated with rational solutions of \PIV\ (\ref{eq:PIV}), it was shown that there are additional solutions than might be anticipated from earlier work of Loutsenko \citep{refLout03}, Oblomkov \citep{refOblomkov}, and Veselov \citep{refVeselov}. As mentioned above, a natural question is whether there are any more?

An interesting open question is whether there is a relationship between vortex dynamics and polynomials associated with rational solutions of other soliton equations, such as the \bq\ equation (\ref{eq:bq}) and the NLS equation (\ref{eq:nls}). Since \PIV\ (\ref{eq:PIV}) arises as a similarity reduction of 
the \bq\ and NLS equations, then one can express rational solutions of these equations in terms of the generalized Hermite and generalized Okamoto polynomials. However both the \bq\ and NLS equations also have rational solutions which involve arbitrary parameters, analogous to the \AM\ polynomials which describe rational solutions of the KdV equation (\ref{eq:kdv}). For the NLS equation (\ref{eq:nls}) an extension of the generalized Hermite polynomials, see \citep{refPAC06nls}, and for \bq\ equation (\ref{eq:bq}) these polynomials involve an extension of the generalized Okamoto polynomials, see \citep{refPACbqrats}. These are obtained by replacing the Hermite polynomial $H_m(z)$, which is defined by the generating function
\[\sum_{m=0}^\infty\frac{H_{m}(z)\,\la^m}{m!} =\exp(2\la z-\la^2),\]
by the polynomial $\Phi_{m}(z)$ defined by
\[\sum_{m=0}^\infty\frac{\Phi_{m}(z)\,\la^m}{m!} =\exp\left(2\la z-\la^2 + \sum_{j=3}^\infty\k_j\la^j\right),\]
where $\k_j$, for $j\geq3$, are arbitrary constants in the definitions of the generalized Hermite and generalized Okamoto polynomials. (This is similar to the relationship between the \VY\ and the \AM\ polynomials where the polynomial $\ph_n(z)$ defined in equation (\ref{eq:2.1}) is replaced by the polynomial $\th_n(z)$ defined in equation (\ref{eq:2.5}).) Therefore the structure of rational solutions of the \bq\ equation (\ref{eq:bq}) and the NLS equation (\ref{eq:nls}) is different to those of the KdV equation (\ref{eq:kdv}) and so other vortex dynamics can possibly be described in terms of the associated generalized polynomials.

Another interesting open question is whether there exist polynomial solutions of the equations
\begin{align}\label{eq51}
&\deriv[2]{P}{z}Q-2\Lambda\deriv{P}{z}\deriv{Q}{z}+\Lambda^2P\deriv[2]{Q}{z}=0,\\ \label{eq52}
&\deriv[2]{P}{z}Q-2\Lambda\deriv{P}{z}\deriv{Q}{z}+\Lambda^2P\deriv[2]{Q}{z}+\mu\left(\deriv{P}{z}Q-\Lambda P\deriv{Q}{z}\right)=0,\\ \label{eq53}
&\deriv[2]{P}{z}Q-2\Lambda\deriv{P}{z}\deriv{Q}{z}+\Lambda^2P\deriv[2]{Q}{z}+\mu z\left(\deriv{P}{z}Q-\Lambda P\deriv{Q}{z}\right)+\kappa PQ=0,\end{align}
where $\Lambda\not=1$, $\mu$ and $\kappa$ are constants, which respectively reduce to equations (\ref{eq:svp1}), (\ref{eq:tvp1}), and (\ref{eq:qbf1}), when $\Lambda=1$. Loutsenko \cite{refLout04} has obtained some polynomial solutions of equations (\ref{eq51}) and  (\ref{eq52}) in the case when $\Lambda=2$, which corresponds to the case of vortices with two strengths of unequal magnitudes.

Patterns of points corresponding to the equilibria of point vortices reveal interesting connections to the classical theory of polynomials and to some of the special polynomials that arise in soliton theory. While some of these connections have been discovered, we believe there is much more that awaits to be found given the large number of results which are known in soliton theory.

\section*{Acknowledgements} 
I would like to thank Mark Ablowitz, Hassan Aref, Carl Bender, Darren Crowdy, Bernard Deconinck, Galina Filipuk, Thanasis Fokas, Rod Halburd, Andy Hone, Igor Loutsenko, Elizabeth Mansfield, and Bryn Thomas for their helpful comments and illuminating discussions. I also thank the referee for some constructive comments which improved the manuscript.

\def\AAM{Acta Appl. Math.}
\def\ARMA{Arch. Rat. Mech. Anal.}
\def\bull{Acad. Roy. Belg. Bull. Cl. Sc. (5)}
\def\AC{Acta Crystrallogr.}
\def\AM{Acta Metall.}
\def\ampa{Ann. Mat. Pura Appl. (IV)}
\def\AP{Ann. Phys., Lpz.}
\def\APNY{Ann. Phys., NY}
\def\APP{Ann. Phys., Paris}
\def\BAMS{Bull. Amer. Math.Soc.}
\def\CJP{Can. J. Phys.}
\def\cmp{Commun. Math. Phys.}
\def\CMP{Commun. Math. Phys.}
\def\cpam{Commun. Pure Appl. Math.}
\def\CPAM{Commun. Pure Appl. Math.}
\def\CQG{Classical Quantum Grav.}
\def\crp{C.R. Acad. Sc. Paris}
\def\CSF{Chaos, Solitons \&\ Fractals}
\def\DE{Diff. Eqns.}
\def\DU{Diff. Urav.}
\def\ejam{Europ. J. Appl. Math.}
\def\EJAM{Europ. J. Appl. Math.}
\def\funk{Funkcial. Ekvac.}
\def\FUNK{Funkcial. Ekvac.}
\def\IP{Inverse Problems}
\def\JAMS{J. Amer. Math. Soc.}
\def\JAP{J. Appl. Phys.}
\def\JCP{J. Chem. Phys.}
\def\JDE{J. Diff. Eqns.}
\def\JFM{J. Fluid Mech.}
\def\JJAP{Japan J. Appl. Phys.}
\def\JP{J. Physique}
\def\JPhCh{J. Phys. Chem.}
\def\JMAA{J. Math. Anal. Appl.}
\def\JMMM{J. Magn. Magn. Mater.}
\def\JMP{J. Math. Phys.}
\def\jmp{J. Math. Phys.}
\def\JNMP{J. Nonl. Math. Phys.}
\def\jpa{J. Phys. A: Math. Gen.}
\def\JPA{J. Phys. A: Math. Gen.}
\def\JPB{J. Phys. B: At. Mol. Phys.} 
\def\jpb{J. Phys. B: At. Mol. Opt. Phys.} 
\def\JPC{J. Phys. C: Solid State Phys.} 
\def\JPCM{J. Phys: Condensed Matter} 
\def\JPD{J. Phys. D: Appl. Phys.}
\def\JPE{J. Phys. E: Sci. Instrum.}
\def\JPF{J. Phys. F: Metal Phys.}
\def\JPG{J. Phys. G: Nucl. Phys.} 
\def\jpg{J. Phys. G: Nucl. Part. Phys.} 
\def\JSP{J. Stat. Phys.}
\def\JOSA{J. Opt. Soc. Am.}
\def\JPSJ{J. Phys. Soc. Japan}
\def\JQSRT{J. Quant. Spectrosc. Radiat. Transfer}
\def\LMP{Lett. Math. Phys.}
\def\LNC{Lett. Nuovo Cim.}
\def\NC{Nuovo Cim.}
\def\NIM{Nucl. Instrum. Methods}
\def\NL{Nonlinearity}
\def\NMJ{Nagoya Math. J.}
\def\NP{Nucl. Phys.}
\def\pl{Phys. Lett.}
\def\PL{Phys. Lett.}
\def\PMB{Phys. Med. Biol.}
\def\PR{Phys. Rev.}
\def\PRL{Phys. Rev. Lett.}
\def\PRS{Proc. R. Soc.}
\def\prsl{Proc. R. Soc. Lond. A}
\def\PRSL{Proc. R. Soc. Lond. A}
\def\PS{Phys. Scr.}
\def\PSS{Phys. Status Solidi}
\def\PTRS{Phil. Trans. R. Soc.}
\def\RMP{Rev. Mod. Phys.}
\def\RPP{Rep. Prog. Phys.}
\def\RSI{Rev. Sci. Instrum.}
\def\SAM{Stud. Appl. Math.}
\def\sam{Stud. Appl. Math.}
\def\SSC{Solid State Commun.}
\def\SST{Semicond. Sci. Technol.}
\def\SUST{Supercond. Sci. Technol.}
\def\TMP{Theo. Math. Phys.}
\def\ZP{Z. Phys.}
\def\OUP{O.U.P.} 
\def\CUP{C.U.P.} 

\def\refpp#1#2#3#4#5{\vspace{-0.2cm}
\bibitem{#1} \textsc{\frenchspacing#2}, \textrm{#3}, #4 (#5).}

\def\refjl#1#2#3#4#5#6#7{\vspace{-0.2cm}
\bibitem{#1}\textsc{\frenchspacing#2}, \textrm{#6}, 
\textit{\frenchspacing#3}, {#4}:#5\ (#7).}

\def\refjltoap#1#2#3#4#5#6#7{\vspace{-0.2cm}
\bibitem{#1} \textsc{\frenchspacing#2}, \textrm{#6}, 
\textit{\frenchspacing#3} 
#5 (#7).}

\def\refbk#1#2#3#4#5{\vspace{-0.2cm}
\bibitem{#1} \textsc{\frenchspacing#2}, \textit{#3}, #4, #5.} 

\def\refcf#1#2#3#4#5#6{\vspace{-0.2cm}
\bibitem{#1} \textsc{\frenchspacing#2}, \textrm{#3},
in \textit{#4}, {\frenchspacing#5}, #6.}

\def\fit{\frenchspacing\it}

\def\adele#1#2#3#4{\refcf{#1}{#2}{#3}{\p\ Transcendents, their Asymptotics and Physical Applications}{P.~Winternitz {\rm and} D.~Levi, Editors}{{NATO ASI Series B: Physics}, \textbf{278}, Plenum, New York (1992) pp.~#4}}

\end{document}